\documentclass[9.5pt,conference]{IEEEtran}
\usepackage{amsmath}
\usepackage{amsthm}
\usepackage{cite}
\usepackage{amssymb}
\usepackage{graphicx}
\usepackage{algorithm}
\usepackage{algorithmic}
\usepackage{bm}
\newtheorem{theorem}{Theorem}

\ifCLASSINFOpdf
\else
\fi
\pdfoutput=1

\begin{document}
\title{Dynamic SINR-Guided Iterative Interference Cancellation for ODDM Systems in Doubly Dispersive Channels}
\author{
	Jiasong~Han\textsuperscript{1},~Xuehan~Wang\textsuperscript{1},~Jintao~Wang\textsuperscript{1,2}\\
	\IEEEauthorblockA{
		\textsuperscript{1}Beijing National Research Center for Information Science and Technology (BNRist),\\
		Dept. of Electronic Engineering, Tsinghua University, Beijing, China\\
		\textsuperscript{2}Research Institute of Tsinghua University in Shenzhen, Shenzhen, China\\
		\{hanjs22@mails., wang-xh21@mails., wangjintao@\}tsinghua.edu.cn}
\thanks{This work was supported in part by Tsinghua University-China Mobile Research Institute Joint Innovation Center.}	
}
\maketitle
\begin{abstract} 
Orthogonal delay-Doppler division multiplexing (ODDM) modulation has recently gained significant attention as a promising candidate to promote the communication reliability in high-mobility environments. Low complexity signal detection  is one of the most significant challenges for ODDM over general physical channels, due to the large channel spreading caused by the fractional delay and Doppler shifts. In this paper, we investigate the low-complexity data detection for ODDM system by utilizing iterative interference cancellation. Based on the theoretical analysis of signal to interference plus noise ratio (SINR) during the iteration, a dynamic SINR-guided approach is proposed to provide a better initialization result. Specifically, we analyze the SINR of each time domain sample before initial estimate with consideration of off-grid delay and Doppler shifts. The iteration is then started from the multi-carrier symbol index which has the best SINR. The corresponding interference is then eliminated for other time domain samples while the SINR for symbol awaiting detection is also updated. Based on the updated SINR, the next multi-carrier symbol index is selected for the same processing until all data symbols have been initialized. Finally, we update the SINR synchronously until the end of the initialization. Simulation experiments indicate that our proposed algorithms demonstrate satisfying convergence and error performance while avoiding the huge complexity introduced by full linear minimum mean squared error (LMMSE) initialization.
\end{abstract}

\IEEEpeerreviewmaketitle
\section{Introduction}
The sixth-generation (6G) mobile systems necessitates the provision of ultra-reliable communication in high-mobility environments, including high-speed railways (HSR), unmanned aerial vehicles (UAVs), and low earth orbit (LEO) satellites \cite{6G_general, CNN}. Orthogonal frequency division multiplexing (OFDM)\cite{OFDM}, which has been widely used in 4G, 5G wireless networks, faces challenges in high-mobility scenarios due to intercarrier interference (ICI) caused by Doppler spread, resulting in significant performance degradation \cite{OMP}.

To address this issue, orthogonal time frequency space (OTFS) modulation has been proposed \cite{OTFS}, which offers superior performance in high-mobility environments by modulating information in the delay-Doppler (DD) domain, where the doubly-selective channel response can be modeled as quasi time-invariant. However, the realization of OTFS modulation is impractical, due to the high out-of-band emission (OOBE) brought by the discontinuity and rectangular pulse-shaping. To solve this problem, orthogonal delay-Doppler multiplexing (ODDM) modulation has been introduced \cite{ODDM}. ODDM modulation is characterized by the design of a pulse train that maintains orthogonality with respect to the DD resolutions \cite{multiple_access}. This approach facilitates superior coupling between the modulated signal and the DD channel, demonstrating enhanced performance compared to OTFS.

In principle, DD domain modulated symbols spread across the entire TF domain \cite{OTFS_LMMSE}. This leads to high complexity of conventional detectors including zero forcing (ZF) and minimum mean squared error (MMSE) \cite{ZF_MMSE} due to the increasing dimension of channel matrices. Thus, low-complexity yet powerful detectors are necessary to achieve the desired error performance of the modulation.

To address this challenge, low-complexity iterative algorithms have been proposed including iterative successive interference cancellation (SIC) maximal ratio combining (MRC) \cite{OTFS_MRC} and iterative SIC-linear minimum mean squared error (LMMSE)\cite{OTFS_LMMSE} detection, which can strike a balance between complexity and performance. \cite{ODDM_MRCMMSE} has analyzed the significant role of good initialization in these iterative algorithms, which can break the error floor in high signal-to-noise ratio (SNR) region. Therefore, numerous studies have designed various initialization methods. The most basic methods are all-zero initialization and single-tap MMSE initialization \cite{OTFS_MRC}. While these methods have low computational complexity, they often perform poorly in many systems. An initialization method based on full LMMSE was proposed in \cite{OTFS_LMMSE}. Although \cite{LU} reduced its computational complexity through LU decomposition, the method still remains computationally intensive. An initialization method based on soft SIC-LMMSE was proposed in \cite{ODDM}. However, when dealing with off-grid ODDM systems, the complexity increases significantly due to the strong scattering effect of the channel matrix. This motivates us to seek an initialization method that offers a favorable balance between computational complexity and error performance.

These pioneering works provide the foundation for our dynamic SINR-guided initialization method, which achieves better performance in these iterative algorithms. In this paper, We analyze the SINR of each time-domain sample before the initial estimate, taking into account off-grid delay and Doppler shifts. For each multi-carrier symbol index, we select the worst SINR among the $N$ symbols as the basis for decision-making. The iteration then begins with the multi-carrier symbol index that has the highest SINR. The corresponding interference is eliminated for other time-domain samples, and the SINR for the symbols awaiting detection is updated accordingly. Based on the updated SINR, the next multi-carrier symbol index is selected for the same processing. This process continues until all data symbols have been initialized. Our research demonstrates that iterative detection algorithms using our proposed initialization method achieves the performance of those using full LMMSE initialization, while maintaining comparable computation complexity with those using all-zero initialization.
  
\textit{Notations}: $\mathbf{A}$, $\mathbf{a}$, $a$ denote a matrix, column vector and scalar, respectively. $\mathbf{A}^H$ and $\mathbf{A}^{-1}$ are its conjugate transposition and inverse. $||\mathbf{A}||$ denotes the norm of $\mathbf{A}$. $[k]_M$ is the modulo operation, and $\lfloor \cdot \rfloor$ is the floor function. $\mathbf{F}_N$ denotes the normalized $N$-point discrete Fourier transform (DFT) matrix.

\section{System Model}

In this section, we first discuss the general framework of ODDM signaling. Subsequently, we present the time-domain input-output (IO) relationship for Zero-Padding (ZP) ODDM, formulated in both discrete and matrix representations.

\subsection{ODDM}

In this paper, we consider ODDM systems over general physical channels operating with a sampling period $T_s$ and a frame duration of $MNT_s$, where each ODDM frame consists of $M$ multi-carrier symbols and each symbol has $N$ subcarriers. An ODDM frame carries $MN$ digital symbols $\{{X[m,n]}|m=0,1, \cdots, M-1, n=0,1, \cdots, N-1\}$ in the delay-Doppler domain, where $X[m,n]$ indicates the data component at the $n$-th subcarrier of the $m$-th symbol. At the transmitter, the inverse discrete Fourier transform (IDFT) is performed on $X[m,n]$ to obtain $\dot{n}$-th sample within the $m$-th symbol as

\begin{align}
    \dot{x}[m,\dot{n}] = \frac{1}{\sqrt{N}} \sum\limits_{n=0}^{N-1} X[m,n] e^{j2\pi \frac{n\dot{n}}{N}}, \quad 0 \leq \dot{n} \leq N-1.
\end{align}

$\dot{x}[m,\dot{n}]$ is then vectorized to obtain the time domain samples as

\begin{align}
    s[k] = \dot{x}\left[ [k]_M, \left\lfloor \frac{k}{M} \right\rfloor \right], \quad 0 \leq k \leq MN - 1.
\end{align}

Then, as mentioned in \cite{ODDM_general}, the continuous time baseband signal can be obtained by applying the sample-wise pulse shaping as

\begin{align}
    s(t) = \sum\limits_{k=0}^{MN-1} s[k] a(t-kT_s),
\end{align}

\noindent where the subpulse $a(t)$ is a square-root Nyquist pulse parameterized by its zero-ISI interval $T_s$ and its duration \( T_a = 2QT_s\), where $Q$ is a positive integer and \( 2Q \ll M\).

According to the sparsity of the delay-Doppler domain channel, the channel can be modeled as

\begin{align}
    h(\tau, t) = \sum\limits_{p=1}^P \rho_p \delta(\tau - \tau_p) e^{j 2\pi \nu_p (t - \tau_p)},
\end{align}

\noindent where \( P \) is the number of propagation paths, \( \rho_p \), \( \tau_p \), and \( \nu_p \) represent the complex gain, delay, and Doppler shift associated with the \( p \)-th path, and \( \delta(\cdot) \) denotes the Dirac delta function. The baseband received signal can be written as

\begin{align}
    r(t) = \sum_{k=0}^{MN-1} s[k] \int_{\tau} a(t - kT_s - \tau) h(\tau, t) d\tau + z(t),
\end{align}

\noindent where $z(t)$ denotes the additive noise. At the receiver, the sample-wise matched filtering (MF) with the impulse response $a(t)$ is applied. The MF output can be derived as

\begin{align}
    r_{\text{MF}}(t) = r(t) * a^*(-t) = \int_{\tau'} r(t - \tau') a^*(-\tau') d\tau'.
\end{align}

It is then sampled at $t = kT_s$ to obtain the digital samples as

\begin{align}\label{eq:io}
    & \nonumber r[k]
    = r_{MF}(t)|_{t=kT_s} \\
    =& \sum\limits_{d=0}^D \sum\limits_{p=1}^P \rho_p g(dT_s - \tau_p)e^{j2\pi \nu_p (kT_s-\tau_p)} s[[k-d]_{MN}] + z[k],
\end{align}

\noindent where $D$ is the total number of delay taps of the equivalent sampled channel and $g(\tau) \triangleq a(\tau) * a^*(-\tau)$ \cite{ODDM_general}.
 
\subsection{Time Domain IO Relation for ZP-ODDM}

\begin{figure}[h!]
    \centering
    \includegraphics[width=1\linewidth]{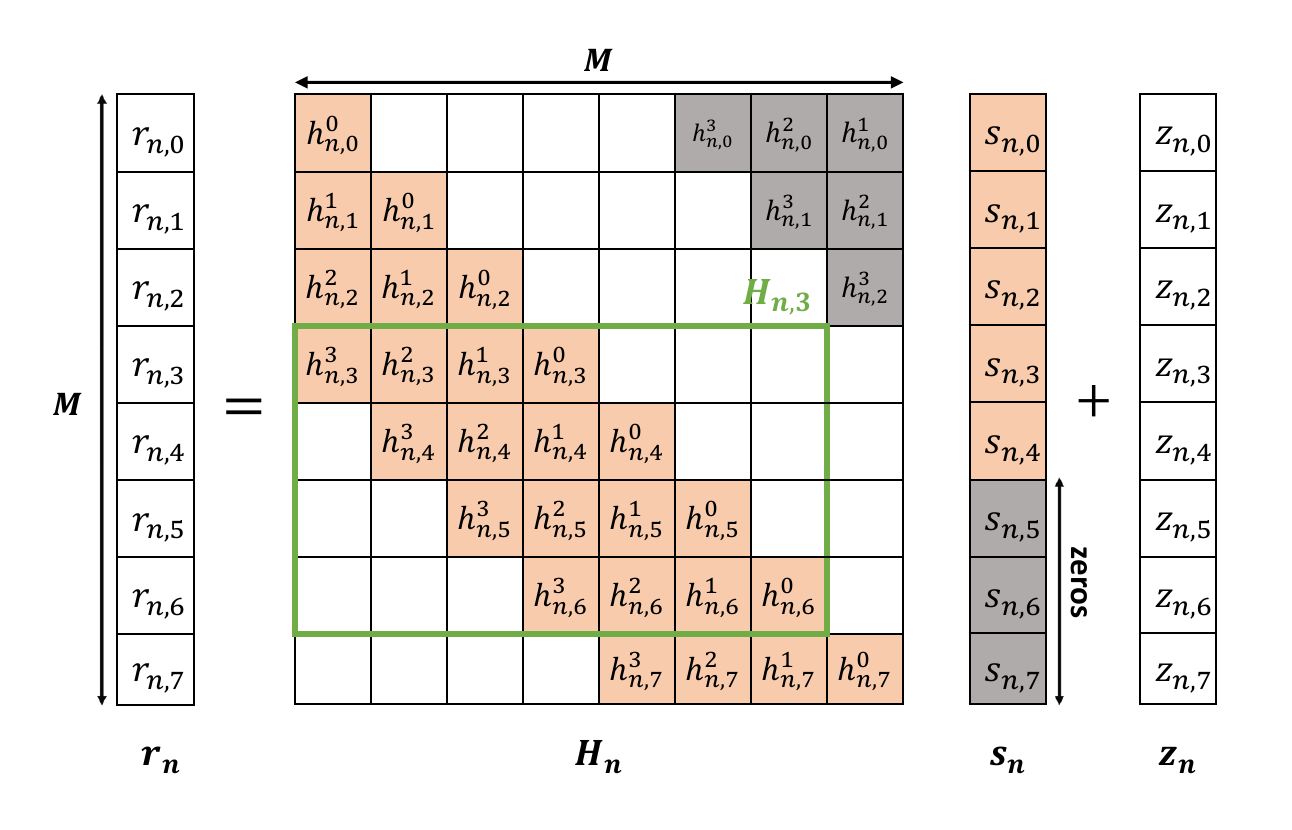}
    \caption{Time domain IO relation for the $n$-th block with $D=3$}
    \label{fig:H_matrix}
\end{figure}

In this paper, we adopt the ODDM system including zero padding as ZP-ODDM. Adding a ZP along the delay dimension in the ODDM delay-Doppler grid can mitigate the interference, which contributes to reduction in detector complexity \cite{OTFS_MRC}. In this scheme, the channel matrix will be divided into $N$ non-interfering parts with the same format. For each part, we only need to detect the first $M'=M-D$ symbols. For ease of illustration, let $l_p \triangleq \frac{\tau_p}{T_s}$ and $k_p \triangleq \nu_p MNT_s$ denote the normalized delay and Doppler shifts, which are not necessarily integer. Then the block-wise IO relation of \eqref{eq:io} can be expressed as

\begin{align}
    &\nonumber r_{n,m} = \sum\limits_{d=0}^{D'} h_{n,m}^d s_{n,m-d} + z_{n,m},
\end{align}

\noindent where we have

\begin{align}
    & h_{n,m}^d = \sum\limits_{p=1}^P \rho_p g((d-l_p)T_s)e^{j\frac{2\pi}{MN} k_p(m-l_p)} e^{j2\pi \frac{nk_p}{N}}, \\
    & D' = \min{(m,D)}.
\end{align}

The block-wise time domain IO relation can be further expressed in a matrix form as shown in Fig. \ref{fig:H_matrix} given by 

\begin{align}
    \mathbf{r}_n = \mathbf{H}_n \mathbf{s}_n + \mathbf{z}_n,
\end{align}

\noindent where $\mathbf{r}_n = [r_{n,0}, \ldots, r_{n,(M-1)}]^T \in \mathbb{C}^{M \times 1}$ is the $n$-th received signal block that contains $M$ elements and each $r_{n,m}$ represents the $m$-th index in the $n$-th block. $\mathbf{s}_n = [s_{n,0}, \ldots, s_{n,(M-1)}]^T \in \mathbb{C}^{M \times 1}$ is the $n$-th transmitted signal block and each $s_{n,m}$ represents the $m$-th element in the $n$-th transmitted block. $\mathbf{H}_n \in \mathbb{C}^{M \times M}$ is the channel matrix for the $n$-th transmitted signal block, and $\mathbf{z}_n$ is the noise vector with variance $\sigma_w^2$.

\section{Proposed Dynamic SINR-Guided Initialization Algorithm}

This section proposes a dynamic SINR-guided initialization algorithm for ODDM systems. We analyze the SINR under perfect SIC and introduces an iterative initialization method that dynamically updates SINR and performs interference cancellation. The complexity of the proposed scheme can be treated as the linear order while an excellent BER performance can be expected, which is illustrated in detail in Section IV.

\subsection{Iterative Interference Cancellation Algorithm}

Let $\overline{\mathbf{r}}_{n,m} = [r_{n,m}, r_{n,m+1}, \cdots, r_{n,m+D}]^T$ denote the received signal vector that corresponds to the signal layer $s_{n,m}$, $\mathbf{H}_{n,m}$ denote the sub-channel for the signal layer $s_{n,m}$, and $\mathbf{h}_{n,m} = [h_{n,m}^0, h_{n,m+1}^1, \cdots, h_{n,m+D}^D]^T$ denote the vector in the channel matrix which acts directly with $s_{n,m}$, respectively. In general, the received signal vector $\overline{\mathbf{r}}_{n,m}$ can be expressed as

\begin{align}
    \overline{\mathbf{r}}_{n,m} = \sum\limits_{j=0}^{D'+D} \mathbf{H}_{n,m}[:,j] s_{n,m'+j} + \mathbf{z}_{n,m},
\end{align}

\noindent where $m' = \max{(m-D,0)}, \; D'=\min{(m,D)}$.

Let \( \overline{s}_{n,m}^{(i)} \) denote the hard estimate of time domain variable detected in the $i$-th iteration. The hard interference cancellation in the $i$-th iteration can then be written as

\begin{align}
    &\nonumber \widetilde{\mathbf{r}}_{n,m}^{(i)} =  \overline{\mathbf{r}}_{n,m} - \sum\limits_{j=0}^{D'-1} \mathbf{H}_{n,m}[:,j]\overline{s}_{n,(m'+j)}^{(i)} \\ & - \sum\limits_{k=D'+1}^{D'+D}\mathbf{H}_{n,m}[:,k]\overline{s}_{n,(m'+k)}^{(i-1)}.
\end{align}

We put the SIC symbols through MRC/LMMSE filter, whose detailed calculation process can be found in \cite{OTFS_MRC, OTFS_LMMSE}. The estimated output sample $\hat{s}_{n,m}^{(i)}$ can be written as

\begin{align}
    \hat{s}_{n,m}^{(i)} = \mathbf{w}_{n,m} \widetilde{\mathbf{r}}_{n,m}^{(i)},
\end{align}

\noindent where the filter \( \mathbf{w}_{n,m} \) can be represented as

\begin{align}
    \mathbf{w}_{n,m}^{MRC} = (\mathbf{h}_{n,m}^H \mathbf{h}_{n,m})^{-1} \mathbf{h}_{n,m}^H, 
\end{align}

\begin{align}
    \mathbf{w}_{n,m}^{MMSE} = (P_t \mathbf{h}_{n,m}^H \mathbf{h}_{n,m} + \sigma_w^2)^{-1} \mathbf{h}_{n,m}^H,
\end{align}

\noindent in MRC and LMMSE, respectively.

Let $\hat{\mathbf{s}}_m^{(i)} = [\hat{s}_{0,m}^{(i)}, \hat{s}_{1,m}^{(i)}, \cdots, \hat{s}_{(N-1)),m}^{(i)}]$ denote the time domain estimated vector. After estimating a multi-carrier symbol index, we transform the estimated samples to the DD domain as

\begin{align}
    \widetilde{\mathbf{x}}_m^{(i)} = \mathbf{F}_N \hat{\mathbf{s}}_m^{(i)}.
\end{align}

The DD domain symbols can be used for hard decision given by

\begin{align}
    \hat{\mathbf{x}}_m^{(i)}[n] = \arg\min\limits_{a \in \mathcal{Q}} |a - \widetilde{\mathbf{x}}_m^{(i)}[n]|,
\end{align}

\noindent which are then converted back to the time domain for the next iteration of SIC as

\begin{align}
    \overline{\mathbf{s}}_m^{(i)} = \mathbf{F}_N^H \hat{\mathbf{x}}_m^{(i)}.
\end{align}

\subsection{SINR-Guided Initialization}

The aforementioned iterative algorithm sequentially estimates the results for each multi-carrier symbol index and utilizes these estimates to perform interference cancellation. This implies that incorrect decisions made in earlier stages may affect subsequent judgments, causing error propagation. As a result, the initialization is especially essential, which motivates us to propose maintaining a dynamically updated SINR array, with each element representing the minimum SINR across $N$ symbols for each multi-carrier symbol index. Using this real-time updated SINR array, we can initialize the multi-carrier symbol index with the highest SINR from beginning to end. To achieve this target, we first present an analysis of the SINR. We define the indicator function as

\begin{align}
    \epsilon_m = \begin{cases}
        1 & \text{if } |\overline{\mathbf{s}}_m| \neq 0, \\
        0 & \text{else},
    \end{cases} 
\end{align}

\noindent which means the $m$-th multi-carrier symbol index has been initialized when $\epsilon_m = 1$. We can use the initialized symbols to do interference cancellation\footnote{Here, we assume perfect interference cancellation, which means that $\overline{\mathbf{s}}_m = \mathbf{s}_m$ when $\epsilon_m = 1$.} as

\begin{align}
    &\nonumber \widetilde{\mathbf{r}}_{n,m} =  \mathbf{H}_{n,m}[:,D']s_{n,m} \\ \nonumber & + \sum\limits_{j=0}^{D'-1} \mathbf{H}_{n,m}[:,j] (1-\epsilon_{m'+j}) \overline{s}_{n,(m'+j)}^{(i)} \\ & + \sum\limits_{k=D'+1}^{D'+D}\mathbf{H}_{n,m}[:,k] (1-\epsilon_{m'+k}) \overline{s}_{n,(m'+k)}^{(i-1)} + \mathbf{z}_{n,m}.
\end{align}

Let $\bm{\varphi}_{n,m} = \mathbf{H}_{n,m}[:,D']s_{n,m}$ represents signal component while $\mathbf{\hat{z}}_{n,m} = \sum\limits_{j=0}^{D'-1} \mathbf{H}_{n,m}[:,j] (1-\epsilon_{m'+j}) \overline{s}_{n,(m'+j)}^{(i)} + \sum\limits_{k=D'+1}^{D'+D}\mathbf{H}_{n,m}[:,k] (1-\epsilon_{m'+k}) \overline{s}_{n,(m'+k)}^{(i-1)} + \mathbf{z}_{n,m}$ represents noise plus interference component. Here, we define the transmit SNR as $\rm{SNR} \triangleq \frac{P_t}{\sigma_w^2}$ with $P_t = \mathbb{E}[|s[k]|^2]$ being the transmit signal power. We have the following theorem that calculates the SINR of symbol $s_{n,m}$, which is defined as $\Pi_{n,m}$.
\begin{theorem}\label{pro}

    \rm{Given the IO relationship of the channel and the SNR, when some of the multi-carrier symbol indices have already been estimated, the SINR of the $(n, m)$-th symbol can be expressed as}

    \begin{align}\label{eq:sinr}
        \nonumber & \Pi_{n,m} = \frac{\mathbb{E}[\bm{\varphi}_{n,m}^H \bm{\varphi}_{n,m}]}{\mathbb{E}[\mathbf{\hat{z}}_{n,m}^H \mathbf{\hat{z}}_{n,m}]} \\
        =& \frac{P_t||\mathbf{H}_{n,m}[:,D']||^2}{\sum\limits_{j=0,j \neq D'}^{D'+D} (1-\epsilon_{m'+j}) P_t ||\mathbf{H}_{n,m}[:,j]||^2 + (D+1) \sigma_w^2}.
    \end{align}
\end{theorem}

\begin{IEEEproof}
    The proof of (\ref{eq:sinr}) is provided in Appendix \ref{th1_proof_apendix}.
\end{IEEEproof}

Based on the analysis above, the overall processing is presented in Algorithm \ref{alg:1}. We first calculate the SINR for all $MN$ symbols. Then, for each multi-carrier symbol index, we select the worst SINR among the $N$ symbols, denoted as $\Phi_m$, as the basis for decision-making. The initialization starts from the multi-carrier symbol index which has the best SINR, which means $\max{(\Phi_m)}$. After estimating the $m$-th multi-carrier symbol index's symbols, we perform a hard decision in the DD domain and then transfer them back to the time domain to update SINR. It could be found that the $m'$-th to the $(m'+D'+D)$-th multi-carrier symbol index SINR need to be updated. Based on the updated SINR, the next multi-carrier symbol index is selected for the same processing until all data symbols have been initialized. During the initialization process of each multi-carrier symbol index, we use SIC-LMMSE algorithm to obtain the estimated results.

\begin{algorithm}[t]
	\renewcommand{\algorithmicrequire}{\textbf{Input:}}
	\renewcommand{\algorithmicensure}{\textbf{Output:}}
	\caption{Dynamic SINR-Guided Initialization for SIC-MRC/SIC-LMMSE}
	\label{alg:1}
	\begin{algorithmic}[1]
		\REQUIRE
		$\mathbf{r}$, $\mathbf{H}$, $\hat{\mathbf{s}}_m = \mathbf{0}_N$, $\epsilon_m = 0$ $\forall m = 0, \cdots , M-1$.
        \STATE \textbf{Initialization:}
        \FOR{$m=0:M'-1$}
            \FOR{$n=0:N-1$}
                \STATE \textbf{Calculate the SINR $\Pi_{n,m}$ according to (\ref{eq:sinr})}  
            \ENDFOR
            \STATE $\Phi_m = \min{(\Pi_{:,m})}$
        \ENDFOR
        \WHILE{$\sum\limits_{m=0}^{M'-1} \epsilon_m \neq M'$}
            \STATE $m = \arg\max\limits_m{(\Phi_m)}$
            \STATE $\epsilon_m = 1$
            \FOR{$n=0:N-1$}
                \STATE \textbf{Estimate $\hat{s}_{n,m}^{(0)}$ using SIC-LMMSE}
                \FOR{$j=1:D+D'$}
                    \STATE \textbf{Update the SINR $\Pi_{n,m'+j}$ according to (\ref{eq:sinr})}
                \ENDFOR
            \ENDFOR
            \FOR{$j=1:D+D'$}
                \STATE $\Phi_{m'+j} = \min{(\Pi_{:,m'+j})}$
            \ENDFOR
        \ENDWHILE
        \STATE $\widetilde{\mathbf{x}}_m^{(0)} = \mathbf{F}_N \hat{\mathbf{s}}_m^{(0)}$
        \STATE $\hat{\mathbf{x}}_m^{(0)}[n] = \arg\min\limits_{a \in \mathcal{Q}} |a - \widetilde{\mathbf{x}}_m^{(0)}[n]|$
        \STATE $\overline{\mathbf{s}}_m^{(0)} = \mathbf{F}_N^H \hat{\mathbf{x}}_m^{(0)}$
        \STATE \textbf{Signal Detection:}
        \STATE \textbf{SIC-MRC/SIC-LMMSE Alogrithm}
		\STATE \textbf{Return} $\hat{\mathbf{x}}.$
	\end{algorithmic}		
\end{algorithm}

\subsection{Complexity Analysis}

Before initialization, the SINR of $M'N$ symbols needs to be calculated separately, with a computational complexity of $\mathcal{O}(M'ND^2)$. During the initialization process, the SINR of $ND$ symbols needs to be updated each time, and such operations need to be performed $M'$ times, resulting in a computational complexity of $\mathcal{O}(M'ND^2)$ for this part. After initialization, hard decision-making requires performing $N$-point FFT/IFFT transformations, with a computational complexity of $\mathcal{O}(M'N\log N)$. The computational complexity of SIC-MRC and SIC-LMMSE is $\mathcal{O}(M'ND)$, as shown in \cite{OTFS_MRC, OTFS_LMMSE}. The overall computational complexity of our proposed dynamic SINR-guided initialization (DSGI) algorithm is $\mathcal{O}(M'ND^2)$. The full LMMSE initialization (FMI) requires the matrix inversion, which leads to a computational complexity of $\mathcal{O}((M')^3N)$. The all-zero initialization (AZI) has a complexity of $\mathcal{O}(M'ND)$. The complexity comparison between the proposed detectors and some benchmark detectors are summarized in TABLE \ref{tab:1}.

\begin{table}[h!]
    \centering
    \caption{COMPLEXITY OF DIFFERENT DETECTION ALGORITHMS}
    \renewcommand\arraystretch{1.5}
    \begin{tabular}{|p{15em}|p{10em}|}
        \hline
        AZI-SIC-MRC/LMMSE & $\mathcal{O}(M'ND)$ \\
        \hline
        DSGI-SIC-MRC/LMMSE & $\mathcal{O}(M'ND^2)$ \\
        \hline
        FMI-SIC-MRC/LMMSE & $\mathcal{O}((M')^3N)$ \\
        \hline
    \end{tabular}
    \label{tab:1}
\end{table}
  
\section{Simulation Results}

In this section, we present the numerical results of the proposed signal detection algorithm. The simulation parameters are shown in TABLE \ref{simulation_para_table}. The delay-power profile is characterized by employing the Tapped Delay Line-B (TDL-B) model according to \cite{3gpp_ts_25_221}. Zero-padding length is set to 32, which is longer than the maximum delay spread of the TDL-B model. In the simulation, we mainly employed three initialization methods including all-zero initialization (AZI), full LMMSE initialization (FMI), and the initialization method we proposed (DSGI). AZI-SIC-MRC/LMMSE and FMI-SIC-MRC/LMMSE are presented as the baseline which have low complexity and high performance, respectively.

\begin{table}[h!]
	\caption{Simulation Parameters}
	\centering
	\label{simulation_para_table}
	\renewcommand\arraystretch{1.2}
	\begin{tabular}{p{15em}p{10em}}
		\hline
		Parameter &
		Typical value\\
		\hline
		Carrier frequency ($f_{c}$)& $4$ GHz\\
		Subcarrier spacing ($\Delta f$)& $15$ kHz\\
		Number of subcarriers ($M$)& 256\\
		Number of ODDM symbols ($N$)& 64\\
		Maximum speed & $1000$ km/h\\
		Delay-power profile& TDL-B\\
		Zero-padding length& 32\\
		Mapping alphabet & 16QAM/64QAM\\
		\hline
	\end{tabular}
\end{table}

\begin{figure}[h!]
    \centering
    \includegraphics[width=0.9\linewidth]{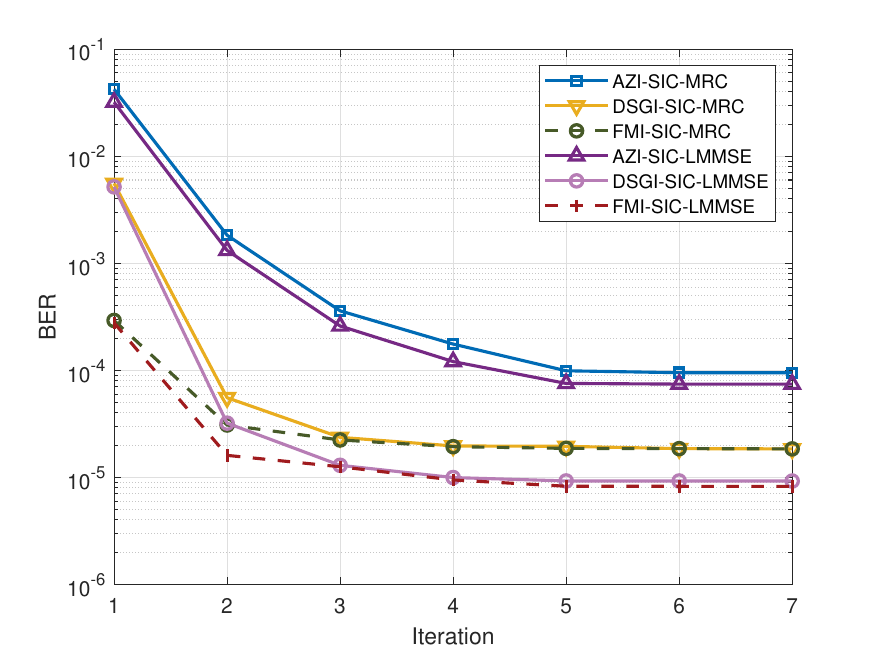}
    \caption{BER against the number of iterations under $\rm{SNR}=24dB$ and 16QAM alphabets.}
    \label{fig:iter}
\end{figure}

The convergence of the proposed scheme is first demonstrated in Fig. \ref{fig:iter} by plotting BER against the number of iterations. In the first iteration, FMI outperforms DSGI. However, after the second iteration, the difference between them becomes virtually indistinguishable. At steady-state convergence, the BER of both FMI and DSGI is reduced by a factor of ten relative to that of AZI. We can see that DSGI initialization may not necessarily yield the most accurate estimate, but it positions the initial point in a favorable location, thereby ultimately achieving a desirable convergence effect.

\begin{figure}[h!]
    \centering
    \includegraphics[width=0.9\linewidth]{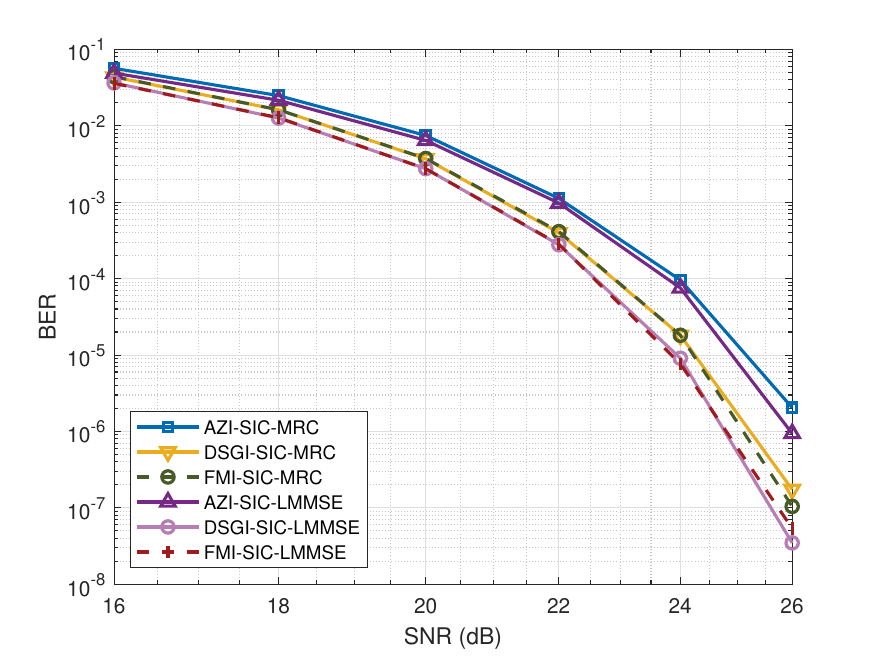}
    \caption{BER against SNR under 16QAM alphabets.}
    \label{fig:16QAM}
\end{figure}

\begin{figure}[h!]
    \centering
    \includegraphics[width=0.9\linewidth]{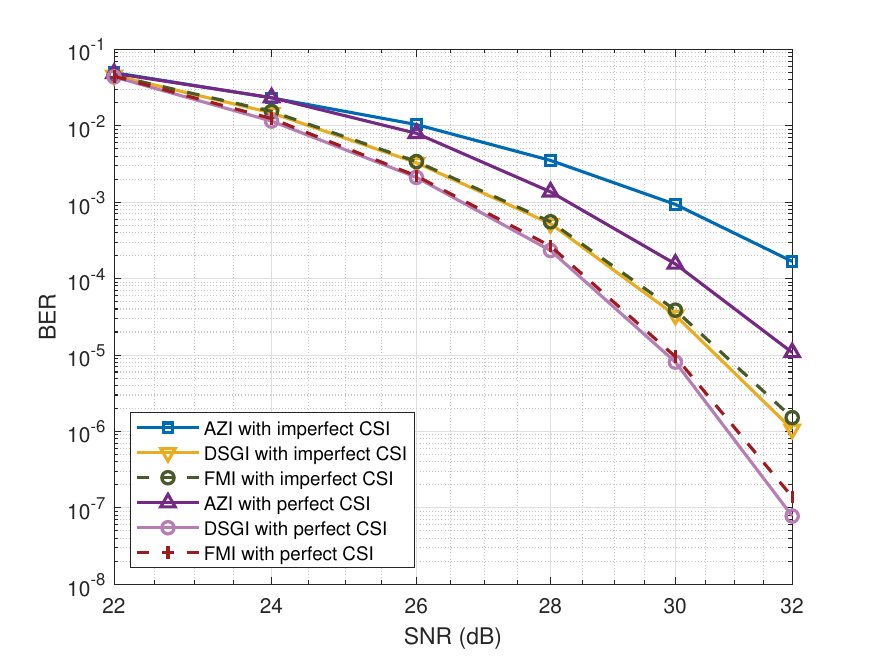}
    \caption{BER against SNR under 64QAM alphabets and SIC-LMMSE detector.}
    \label{fig:64QAM}
\end{figure}

\begin{figure*}[t]
	\begin{equation}
		\small
		\begin{aligned}
			\mathbb{E}[\hat{\mathbf{z}}_{n,m}^H \hat{\mathbf{z}}_{n,m}] &= \mathbb{E}[(\sum\limits_{j=0,j \neq D'}^{D'+D} \mathbf{H}_{n,m}[:,j] (1-\epsilon_{m'+j}) \overline{s}_{n,(m'+j)} + \mathbf{z}_{n,m})^H(\sum\limits_{j=0,j \neq D'}^{D'+D} \mathbf{H}_{n,m}[:,j] (1-\epsilon_{m'+j}) \overline{s}_{n,(m'+j)} + \mathbf{z}_{n,m})] \\ 
            &= \sum\limits_{i=0,i \neq D'}^{D'+D} \sum\limits_{j=0,j \neq D'}^{D'+D} (1-\epsilon_{m'+i})(1-\epsilon_{m'+j})(\mathbf{H}_{n,m}[:,i])^H(\mathbf{H}_{n,m}[:,j]) \mathbb{E}[s_{n,m-i} s_{n,m-j}^*] + (D+1)\sigma_w^2 \\
            &= \sum\limits_{j=0,j \neq D'}^{D'+D} (1-\epsilon_{m'+j}) P_t ||\mathbf{H}_{n,m}[:,j]||^2 + (D+1) \sigma_w^2.
		\end{aligned}
		\label{eq:prove_noise}
	\end{equation}
	\hrulefill
\end{figure*}

Fig. \ref{fig:16QAM} illustrates the BER performance under 16QAM against SNR. We have observed that regardless of the initialization method employed, the performance of SIC-LMMSE consistently surpasses that of SIC-MRC. Our proposed DSGI initialization method demonstrates comparable performance to the FMI initialization, with the SNR gain of 1 dB over AZI at $\rm{BER}=10^{-5}$. As the SNR increases, both FMI and our proposed DSGI exhibit growing performance gains over AZI. 

Fig. \ref{fig:64QAM} illustrates BER under 64QAM modulation and SIC-LMMSE detector. This figure primarily compares the performance of different detectors under perfect Channel State Information (CSI) and imperfect CSI , with the imperfect CSI being set at a Normalized Mean Square Error (NMSE) of $-10\rm{dB}$ ($\rm{NMSE} = \frac{\sum_{p=1}^P |\hat{h}_p - h_p|^2}{\sum_{p=1}^P |h_p|^2}$). It can be observed that DSGI and FMI still maintain good performance under imperfect CSI. This indicates that the proposed method exhibits good robustness against channel estimation errors.

\section{Conclusion}
This paper explored the iterative interference cancellation-based data detection in ODDM systems over general off-grid physical channels. We first derived the input-output relationship for the ZP-ODDM system over general off-grid physical channels. Then we analyze the SINR. Based on the analysis, we introduced a dynamic SINR-guided interference cancellation algorithm. We set a dynamically updated SINR array. Using this real-time updated SINR array, we can initialize the multi-carrier symbol index with the highest SINR from beginning to end. Our proposed algorithms demonstrate satisfying convergence and error performance while avoiding the huge complexity introduced by full LMMSE initialization.

\appendices
\section{Proof of Theorem \ref{pro}}
\label{th1_proof_apendix}

We define $\rm{SNR} \triangleq \frac{P_t}{\sigma_w^2}$ with $P_t = \mathbb{E}[s[k]^2]$ denoting the transmit signal power. The signal power of $s_{n,m}$ can be calculated as follows

\begin{align}
    \nonumber \mathbb{E}[\bm{\varphi}_{n,m} \bm{\varphi}_{n,m}^*] &= \mathbb{E}[(\mathbf{H}_{n,m}[:,D']s_{n,m})^H (\mathbf{H}_{n,m}[:,D']s_{n,m})] \\  \nonumber  &= P_t ||\mathbf{H}_{n,m}[:,D']||^2.
\end{align}

In the DD domain, symbols are independent. Due to IDFT's unitary property, the samples in the time Domain are also uncorrelated of each other, which means \( \mathbb{E}[s_{n,i}s_{n,j}^*] = P_t \delta[i-j] \), where $\delta[\cdot]$ denotes the Kronecker delta function. The calculation of noise plus interference power is shown in Equation (\ref{eq:prove_noise}) which completes the proof of Theorem \ref{pro}.

\section*{Acknowledgment}
This work was supported by the Beijing Natural Science Foundation under Grant QY25034 and the National Natural Science Foundation of China under Grant 624B2079.

\bibliographystyle{IEEEtran}
\bibliography{ref-sum}

\end{document}